\documentclass[a4paper,10pt,twoside]{cpc-hepnp}
\usepackage{upgreek,fancyhdr}
\usepackage{multicol,color}
\usepackage{graphicx,subfigure,dcolumn}
\usepackage{epsfig}
\usepackage{booktabs,slashed}
\usepackage{amssymb,bm,mathrsfs,bbm,amscd,amsfonts}
\usepackage[tbtags]{amsmath}
\usepackage{lastpage}
\newcommand{\f}{\begin{equation}}
\newcommand{\ff}{\end{equation}}
\newcommand{\fa}{\begin{eqnarray}}
\newcommand{\ffa}{\end{eqnarray}}

\providecommand{\U}[1]{\protect\rule{.1in}{.1in}}
\begin{document}

\title{\boldmath Holographic superconductor on a novel insulator
\thanks{Supported by the Natural Science Foundation of China
under Grant Nos. 11575195, 11775036 and 11305018. Y.L. also
acknowledges the support from Jiangxi young scientists (JingGang
Star) program and 555 talent project of Jiangxi Province. J. P. Wu
is also supported by Natural Science Foundation of Liaoning
Province under Grant No.201602013.}}

\author{Yi Ling  $^{1,5}$ \email{lingy@ihep.ac.cn}
\quad Peng Liu  $^{2}$ \email{phylp@jnu.edu.cn}
\quad Jian-Pin Wu  $^{3}$ \email{jianpinwu@mail.bnu.edu.cn}
\quad Meng-He Wu  $^{4,1}$ \email{menghewu.physik@gmail.com}}

\maketitle

\address{
$^1$ Institute of High Energy Physics, Chinese Academy of Sciences, Beijing 100049, China\ \\
$^2$ Department of Physics, Jinan University, Guangzhou 510632, China\ \\
$^3$ Institute of Gravitation and Cosmology, Department of Physics, School of Mathematics and Physics, Bohai University, Jinzhou 121013, China\ \\
$^4$ Center for Relativistic Astrophysics and High Energy Physics, Department of Physics, School of Sciences, Nanchang University, Nanchang 330031, China\ \\
$^5$ School of Physics, University of Chinese Academy of Sciences, Beijing 100049, China}
\maketitle

\begin{abstract}
We construct a holographic superconductor model, based on
a gravity theory, which exhibits novel metal-insulator transitions.
We investigate the condition for the condensation of the scalar
field over the parameter space, and then focus on the
superconductivity over the insulating phase with a hard gap, which is
supposed to be Mott-like. It turns out that the formation of the
hard gap in the insulating phase benefits the superconductivity. This
phenomenon is analogous to the fact that the pseudogap phase can
promote the pre-pairing of electrons in high $T_c$ cuprates. We
expect that this work can shed light on understanding the
mechanism of high $T_c$ superconductivity from the holographic
side.
\end{abstract}




\section{Introduction}

\label{sec:intro} Mott physics, a mechanism for metal-insulator
transitions due to electron-electron interactions, plays a crucial
role in understanding strongly correlated phenomena in a many-body
system (see
Refs.~\cite{MottReview1,MottReview2,MottReview3,Phillips:2010uy} for
reviews). Typical examples include transition metal oxides like
NiO and CoO, as well as the superconducting cuprates, fullerene
compounds like $C_{60}$ and $C_{70}$, and organic conductors
\cite{MottReview2,MottReview3}. Most Mott materials exhibit a hard
gap rather than a power law behavior of frequency dependence as for the
soft gap, which can be observed in the spectral function of single
particles as well as in the optical conductivity. The
spectral weight transfer and the formation of the hard gap are viewed
as two key features of Mott physics \cite{Phillips:2010uy}. It is
widely believed that the strong electron-electron correlation is
responsible for these characteristics of Mott materials. This means that conventional methods developed using perturbation
techniques are, unfortunately, ineffective. AdS/CFT correspondence, as a
powerful tool which in the large N limit maps a strongly coupled
quantum field theory to a weakly coupled gravitational theory, may
provide a different viewpoint on these strongly correlated systems
in condensed matter physics
\cite{Maldacena:1997re,Gubser:1998bc,Witten:1998qj,Aharony:1999ti}.
Several novel localization
mechanisms have been proposed in recent years using this approach, by constructing
different lattice structures as deformations of the bulk geometry
\cite{Ling:2014saa,Donos:2012js,Donos:2014oha,Donos:2013eha,Donos:2014uba,Gouteraux:2014hca,
Baggioli:2014roa,Kiritsis:2015oxa,Ling:2015epa,Ling:2015exa,Baggioli:2016oju}.
In this way, Peierls insulators \cite{Ling:2014saa},
polaron-localization insulators \cite{Baggioli:2014roa}, and Mott-like
insulators \cite{Ling:2015epa,Ling:2015exa,Baggioli:2016oju} as
well as other novel insulating phases, have been implemented and
many exciting properties have been observed
\cite{Donos:2012js,Donos:2014oha,Donos:2013eha,Donos:2014uba,Gouteraux:2014hca,Kiritsis:2015oxa,Montull:2012fy},
some of which resemble those found in strongly correlated
electronic systems. AdS/CFT duality has provided an intuitive and
geometric scenario for understanding the phenomena in strongly
correlated electronic systems.

It is still challenging to understand the mechanism of high
temperature superconductivity. In contrast to conventional
superconductors, for which the normal state is metallic and the
superconductivity emerges due to the formation of Cooper pairs by
means of electron-phonon interactions, it is found that high
temperature copper-oxide superconductors are strongly correlated
electronic systems, which can be generated by doping a Mott
insulator \cite{MottReview1}. The mechanism of pairing electrons
to form superconductivity in high $T_c$ cuprates is still not
clear, but it has been found that before entering the
superconducting phase, some interesting phases with pseudogap and
competing orders are involved. It is very desirable to explore the
mechanism of high $T_c$ superconductivity using a holographic approach.
Until now, for most holographic models of superconductors in the
literature, the normal state is either a perfect conductor
\cite{Hartnoll:2008vx,Hartnoll:2008kx,Horowitz:2010gk,Cai:2015cya},
a metal \cite{Horowitz:2013jaa} or an ordinary insulator
\cite{Ling:2014laa}. Thus, to mimic high $T_c$ superconductivity
by holography it is essential to construct a holographic model
dual to a Mott insulator at first. As the first step we have
successfully proposed a two-gauge formalism in gravity theory and
built a Mott-like insulator in Ref.~\cite{Ling:2015exa}, which is
characterized by the emergence of a hard gap in the optical
conductivity. This appealing progress stimulates us to further
explore the superconductivity based on this Mott-like insulating
phase in a holographic scenario, which is the main purpose of this
work.

Our paper is organized as follows. In Section
{\ref{sec-setup}}, we present the holographic setup for the
construction of a superconductor in two-gauge formalism with
a Q-lattice structure. Section {\ref{sec-ps}} contains four
subsections where the phase structure, the relation between the
gap in the normal state and the critical temperature, and solutions
to scalar condensation and optical conductivity in the superconducting
phase are studied in detail. The conclusion and discussion are
presented in Section {\ref{sec-dis}}.

\section{Holographic setup} \label{sec-setup}

A holographic model which exhibits a novel metal-insulator
transition with a Q-lattice structure has been investigated in
Ref.~\cite{Ling:2015exa}. To construct a superconductor based on this
model, we introduce an additional charged scalar field with $U(1)$
gauge symmetry into the system such that the  action becomes
\begin{eqnarray}
\label{action-S}
S=S_1+S_2,
\end{eqnarray}
where $S_1$ and $S_2$ are, respectively,
\begin{eqnarray}
&&
\label{action-S1}
S_1=\frac{1}{2\kappa^2}\int
d^4x\sqrt{-g}\Big[R+6
-|\nabla\Phi|^2-m^2|\Phi|^2
-\frac{1}{4}F^2
-\frac{Z(\Phi)}{4}G^2\Big],\,
\\
&&
\label{action-S2}
S_2=\frac{1}{2\kappa^2}\int d^4x\sqrt{-g}\Big(-|D_{\mu}\Psi|^2-M^2|\Psi|^2\Big)\,.
\end{eqnarray}
Notice that we have set the AdS radius $L=1$. The action $S_1$ is
proposed in Ref.~\cite{Ling:2015exa}, which contains background
solutions dual to a Mott-like insulating phase. $\Phi$ is a
neutral complex scalar field with mass $m$ which is
responsible for the breaking of translational symmetry in spatial
directions, thus dubbed the Q-lattices \cite{Donos:2013eha}.
$F=dA$, $G=dB$ are the curvatures of two $U(1)$ gauge fields $A$ and
$B$, respectively. The $B$ field is treated as the Maxwell field and
we concentrate on its transport properties in this paper.
$Z(\Phi)=(1-\beta|\Phi|^2)^2$ with $\beta$ being positive, which
describes the interaction between the Q-lattice $\Phi$ and the
Maxwell field $B$ in bulk geometry. This action with two gauge fields plays a crucial role in obtaining an insulating phase with a hard gap.
  When the hard gap is present, the coupling term $Z\to 0$ on the horizon, which means that the effect of the term
$\frac{Z(\Phi)}{4}G^2$ is not strong enough to deform the IR fixed point from $AdS_2$ geometry, which
is dual to a metallic phase, to a new one dual to an insulating phase. Therefore the
second gauge field term $F^2/4$ is introduced to obtain an insulating phase with a hard gap.
The action $S_2$ supports a
superconducting black brane whenever the $U(1)$ gauge
symmetry associated with $B$ is spontaneously broken
\cite{Hartnoll:2008vx}. $\Psi$ is the charged complex scalar field
with mass $M$, which can be written as $\Psi=\psi e^{i\theta}$
with $\psi$ being a real scalar field and $\theta$ a St\"uckelberg
field. $D_{\mu}=\partial_{\mu}-ieB_{\mu}$ is the covariant
derivative where $e$ is the charge associated with the
Maxwell field $B$. For convenience, we choose the gauge $\theta=0$
and therefore $S_2$ can be rewritten as
\fa
S_2=\frac{1}{2\kappa^2}\int
d^4x\sqrt{-g}\Big[-(\nabla_{\mu}\psi)^2-(M^2+e^2B_{\mu}B^{\mu})\psi^2\Big]\,.
\label{action-S2-v1} \ffa
The equations of motion can be derived
from the actions (\ref{action-S1}) and (\ref{action-S2-v1}) as
\begin{eqnarray}
&&
\label{eom-g}
R_{\mu\nu}-\Big(3+\frac{1}{2}R\Big)g_{\mu\nu}
+\frac{1}{2}(T^A_{\mu\nu}+T^B_{\mu\nu}+T^{\Phi}_{\mu\nu}+T^{\psi}_{\mu\nu})=0\,,
\
\\
&&
\label{eom-F}
\nabla^{\mu}F_{\mu\nu}
=0\,,
\
\\
&&
\label{eom-G}
\nabla^{\mu}\big[(1-\beta|\Phi|^2)^2G_{\mu\nu}\big]
-2e^2B_{\nu}\psi^2
=0\,,
\
\\
&&
\label{eom-Phi}
\Big[\nabla^2-m^2+\frac{1}{2}\beta(1-\beta|\Phi|^2)G^2\Big]\Phi=0
\,,
\
\\
&&
\label{eom-psi}
\big[\nabla^2-(M^2+e^2B^2)\big]\psi=0\,,
\end{eqnarray}
where
\begin{eqnarray}
&&
\label{T-A}
T^A_{\mu\nu}=\frac{1}{4}g_{\mu\nu}F^2-F_{\mu\rho}F_{\nu}^{\ \rho}\,,
\
\\
&&
\label{T-B}
T^B_{\mu\nu}=(1-\beta|\Phi|^2)^2\Big(\frac{1}{4}g_{\mu\nu}G^2-G_{\mu\rho}G_{\nu}^{\ \rho}\Big)\,,
\
\\
&&
\label{T-Phi}
T^{\Phi}_{\mu\nu}=-2\nabla_{(\mu}\Phi\nabla_{\nu)}\Phi^{\ast}+g_{\mu\nu}(|\nabla\Phi|^2+m^2|\Phi|^2)
\,,
\
\\
&&
\label{T3}
T^{\psi}_{\mu\nu}=-2(\nabla_{\mu}\psi\nabla_{\nu}\psi+e^2B_{\mu}B_{\nu}\psi^2)
+g_{\mu\nu}\big[(\nabla_{\mu}\psi)^2+(M^2+e^2B_{\mu}B^{\mu})\psi^2\big]
\,.
\end{eqnarray}

The gravitational dual of the normal phase ($\psi=0$) has
been numerically constructed in Ref.~\cite{Ling:2015exa}, in which the
ansatz is
\fa
ds^2&=&{1\over
z^2}\Big[-(1-z)p(z)Udt^2+\frac{dz^2}{(1-z)p(z)U}+V_1dx^2+V_2dy^2\Big]\,,\nonumber\\
A&=&\mu(1-z)a dt\,,\nonumber\\
B&=&\mu(1-z)b dt\,,\nonumber\\
\Phi&=&e^{i \hat{k}x}z^{3-\Delta}\phi\,,\nonumber\\
\psi&=&0\,, \label{bs} \ffa where $p(z)=1+z+z^2-\mu^2z^3/4$ and
the scaling dimension of the scalar field
$\Delta=3/2+(9/4+m^2)^{1/2}$. All the functions $U,V_1,V_2,a,b$
and $\phi$ depend on the radial coordinate $z$ only. Throughout this
paper we set $m^2=-2$ so that $\Delta=2$. For a given coupling
parameter $\beta$, each black hole solution is specified by four
scaling-invariant parameters, {\it i.e.}, the Hawking temperature
$\hat{T}/\mu$ with $\hat{T}=(12-\mu^2)U(1)/16\pi$, the lattice
amplitude $\hat{\lambda}/\mu^{3-\Delta}$ with
$\hat{\lambda}=\phi(0)$, the wave vector $\hat{k}/\mu$ and
$b_0\equiv b(0)$, where without loss of generality we have set
$a(0)=1$. For convenience, these quantities are abbreviated as
$T$, $\lambda$ and $k$ in what follows.

The phase structure for the normal state has been investigated in detail in
Ref.~\cite{Ling:2015exa}. We briefly review its main
features over the parameter space $(\lambda, k, \beta, b_0)$ in the
low temperature region, which is essential for us to explore the
superconducting phase in the current work. In the ordinary Q-lattice
background $(\lambda, k)$ with $\beta=0$ and a single gauge field, a
novel metal-insulator transition has been found by changing
the parameters $(\lambda, k)$ in Ref.~\cite{Donos:2013eha}. Later, the
specific phase diagram over the $(\lambda, k)$ plane was given in Refs.~\cite{Ling:2014laa} and \cite{Ling:2015dma}. Qualitatively, one
finds that the region with large $k$ and small $\lambda$ falls into the
metallic phase, while the region with small $k$ and large
$\lambda$ falls into the insulating phase. This rule is also
consistent with Mott's thought experiment, since small wave-number
$k$ implies a larger lattice constant such that it becomes harder
for electrons to hop to their neighbor sites, leading to an
insulating phase. However, for the above insulating phases the
hard gap is absent, which has been justified in the plot of
optical conductivity \cite{Donos:2013eha,Ling:2014laa}.
Remarkably, after introducing the interaction term with $\beta$ in
two-gauge formalism, we found in Ref.~\cite{Ling:2015exa} that a
hard gap can eventually emerge with the increase of the parameter
$\beta$ for an insulating phase. Next, we intend to investigate
the key features of superconductivity over such Mott-like
insulating phases with hard gaps.

In the holographic approach, the superconducting phase is achieved by
finding non-trivial solutions of the charged scalar hair. Once its
back reaction to the background is taken into account, we need to
numerically solve the equations of motion
(\ref{eom-g})-(\ref{eom-psi}). Without loss of generality, let us
set $M^2=-2$. Then, beyond any details of equations, it is easy to
see that the asymptotical behavior of $\psi$ at infinity is \fa
\label{asy-psi} \psi=z\psi_1+z^2\psi_2\,. \ffa Here, we treat
$\psi_1$ as the source and $\psi_2$ as the expectation value of
the scalar operator in dual quantum field theory. Since we expect
the condensation will emerge without being sourced, we set
$\psi_1=0$.

\section{Condensation and phase structure}\label{sec-ps}
In this section we demonstrate the main numerical results. Superconductivity phase
structures will be discussed first. After that, a
connection between the critical temperature for condensation
and the formation of a gap in the normal state will be shown. Numerical solutions to the background with scalar
hair are obtained explicitly and it is verified that the phase transition is
second order. Finally, the frequency behavior of
optical conductivity in the superconducting phase is investigated in
detail.

\subsection{Phase structure}

The charged scalar is expected to condensate when the Hawking temperature
of the black brane drops to some critical temperature
$T_c$, which reflects the instability of the AdS background with a
violation of the usual $BF$ bound. The condition for the
occurrence of condensation depends on the charge of the scalar
field as well as the background which is specified by the system
parameters. In this section we will analyze the condition for
condensation and obtain the critical temperature $T_c$ for
different system parameters.

\begin{figure}
\center{
\includegraphics[scale=0.5]{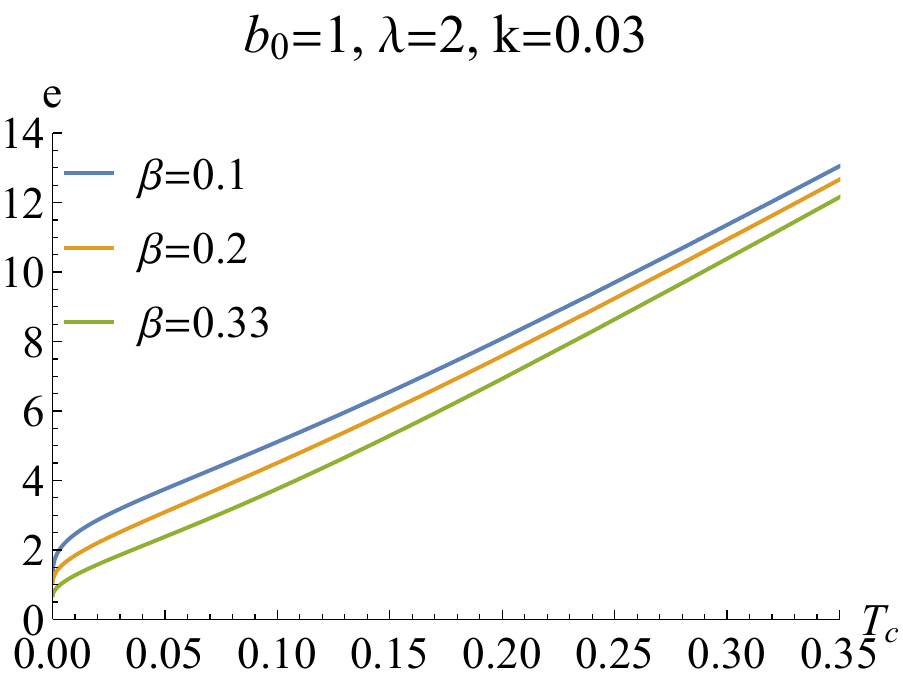}\ \hspace{0.1cm}
\includegraphics[scale=0.5]{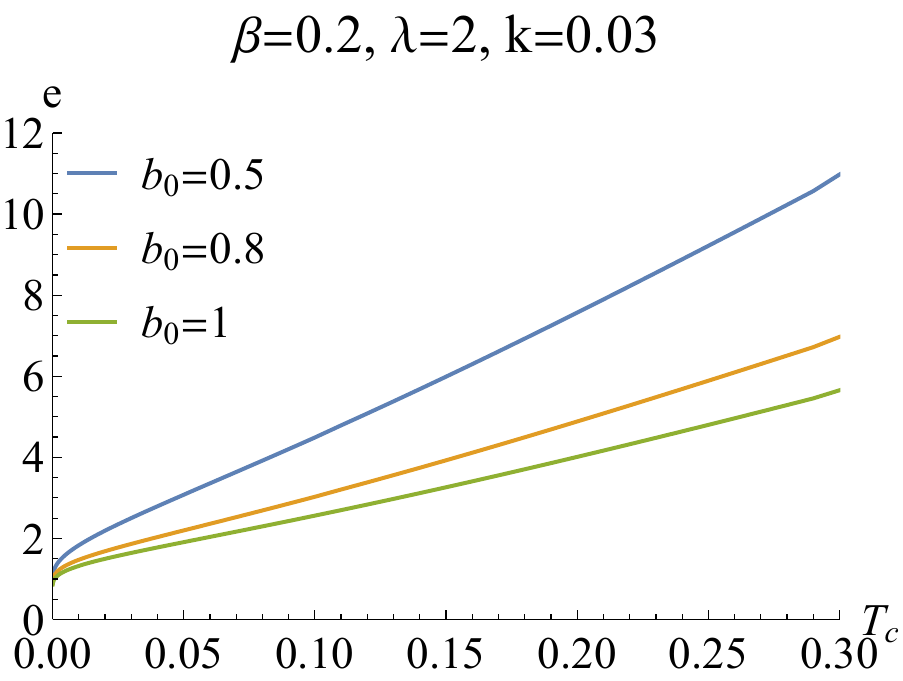}\ \hspace{0.1cm}
\includegraphics[scale=0.5]{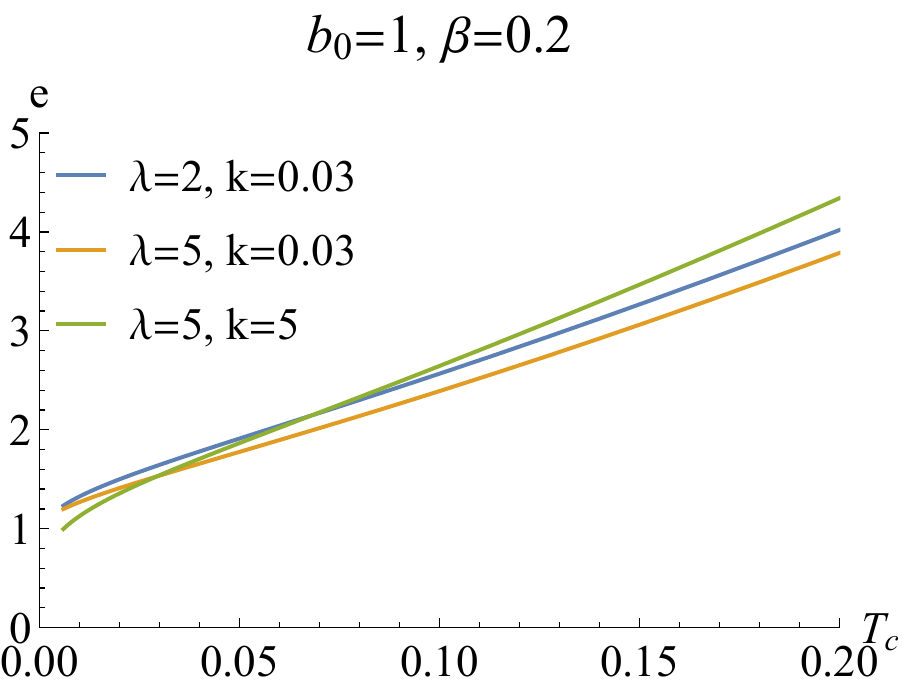}\ \hspace{0.1cm}
\caption{\label{fig-evsTc}
The charge of the scalar field as a function of the critical temperature $T_c$
for some representative values of the system parameters.}}
\end{figure}

\begin{figure}
\center{
\includegraphics[scale=0.5]{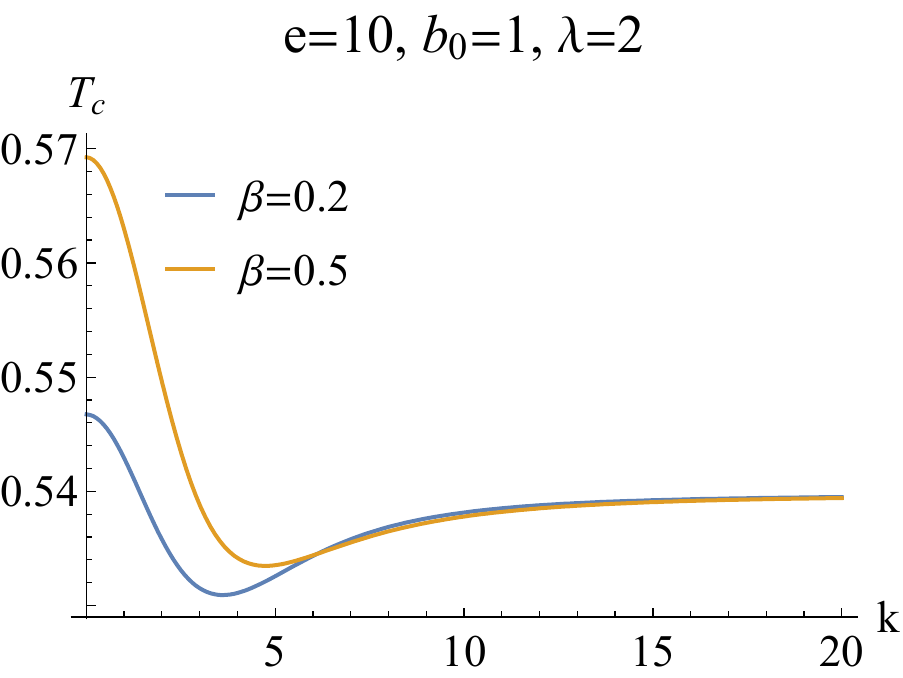}\ \hspace{0.1cm}
\includegraphics[scale=0.5]{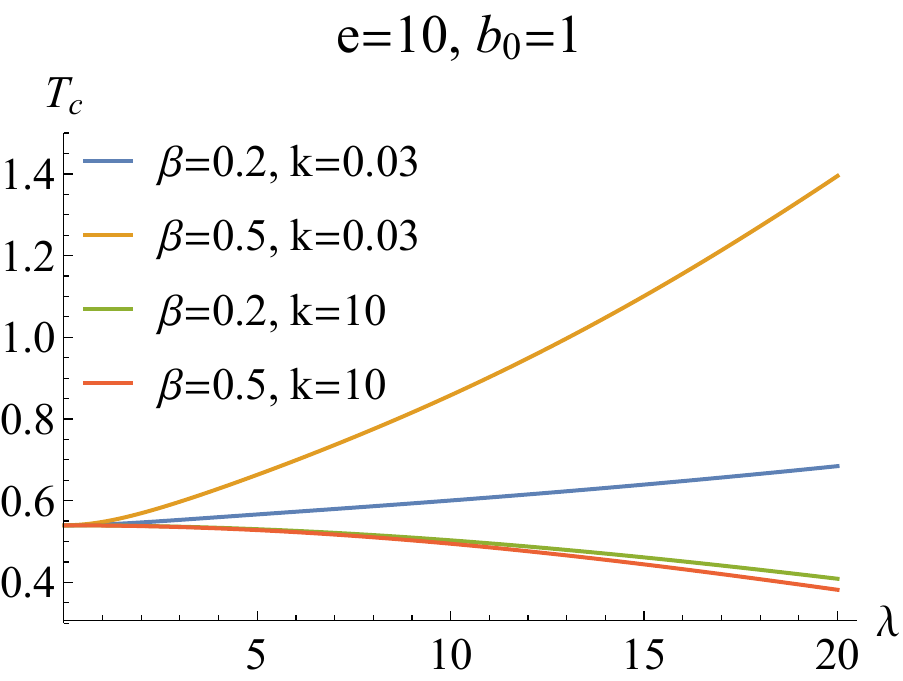}\ \hspace{0.1cm}
\includegraphics[scale=0.4]{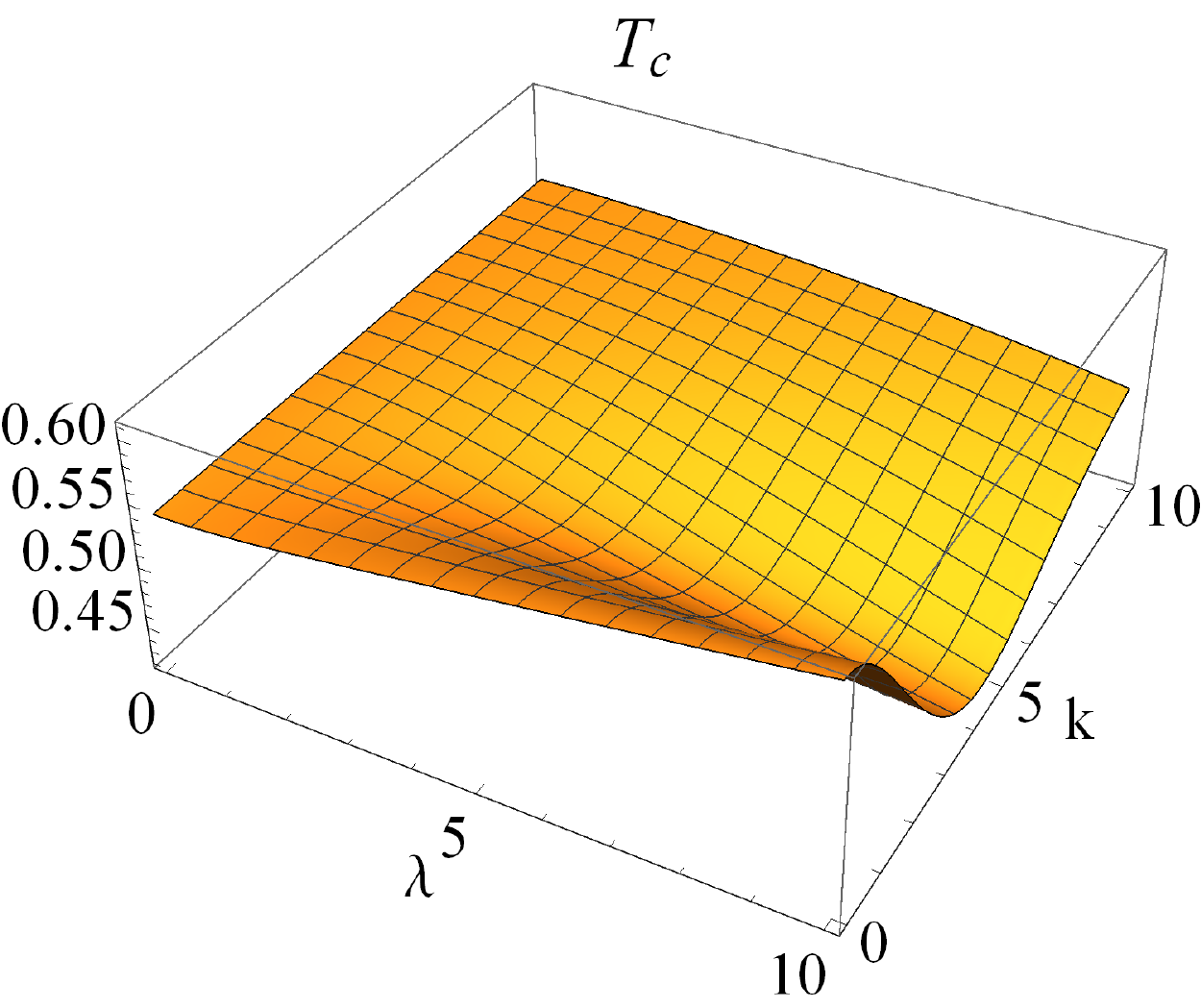}\ \hspace{0.1cm}
\caption{\label{fig-Tc}
The critical temperature $T_c$ as a function of system parameters: (a)
the relation between $T_c$ and $k$ for selected parameters; (b) the relation between $T_c$ and $\lambda$ for selected parameters; and (c)
 a 3D plot of $T_c$ as a function of $\lambda$ and $k$ for $b_0=1$ and $\beta=0.2$.}}
\end{figure}

The critical temperature for the formation of the superconducting
phase can be estimated by finding static normalizable modes of the
charged scalar field on a fixed background (\ref{bs}), which has
been described in detail in Refs.~\cite{Horowitz:2013jaa,Ling:2014laa}.
To this end, one may turn this problem into a positive
self-adjoint eigenvalue problem for $e^2$ and so we rewrite the
equation of motion for the charged scalar field in Eq.~(\ref{eom-psi})
as \fa \label{eom-psi-eigen} -(\nabla^2+2)\psi=-e^2B^2\psi\,, \ffa
which can be numerically solved once a specific metric of the
background is given. Figure  \ref{fig-evsTc} illustrates the
charge of the scalar field as the function of the critical
temperature $T_c$ for some representative values of the system
parameters. First, in general, $T_c$ is
always increasing with the charge $e$. This tendency is the same
as that in other holographic models as described in
Refs.~\cite{Horowitz:2013jaa,Ling:2014laa}. It indicates that the
increase of the charge makes the condensation easier. We shall fix
$e=10$ throughout this paper without loss of generality. Second,
we observe that $T_c$ increases with $\beta$ ($b_0$) when the
other parameters are fixed (Fig.~\ref{fig-evsTc}(a) and (b)) but $T_c$ may be a non-monotonic function
of $\lambda$ or $k$ (Fig.~\ref{fig-evsTc}(c)), which implies there are some interesting
phenomena worth exploring. Thus we plot the critical
temperature $T_c$ as a function of lattice parameters
$(\lambda,k)$ in Fig.~\ref{fig-Tc} and summarize our main
observations as follows.
\begin{itemize}
\item For large $k$, we find $T_c$ rises and approximately
saturates to a constant, which coincides with the phenomenon
observed in the simplest superconductor model on the
Q-lattice \cite{Ling:2014laa}. This is not surprising since in that
region the hard gap is disappearing or the system simply enters a
metallic phase. Also, when $k$ is large the lattice effect is
suppressed, and hence the system is similar to condensation on the AdS-RN background.

One remarkable phenomenon is observed in the small $k$ region
(Fig.~\ref{fig-Tc}(a) and (c)).
 That is, $T_c$ rises as the wavenumber $k$ decreases, with the
other parameters fixed, which is contrary to the tendency observed
in previous holographic superconductor model on the Q-lattice
\cite{Ling:2014laa}, in which the critical temperature always
goes down with the decrease of wavenumber $k$ until it approaches
zero at tiny $k$, implying that no condensation could take place
in such deep insulating phases \cite{Ling:2014laa}. This
difference is significant because it indicates that in
the current setup the superconductivity becomes {\it easier} as the
system enters a deep insulating phase. This
difference results from the involvement of the coupling term with
non-zero $\beta$. In particular, the larger $\beta$ is, the higher
$T_c$ is, as shown in Fig.~\ref{fig-Tc}(a).
Physically, we know that this term is responsible for the
formation of a hard gap when the system falls into an insulating
phase. Therefore, we argue that the presence of a hard
gap in the insulating phase benefits the condensation of the scalar
hair, reminiscent of the phenomenon observed in copper-oxide
superconductors where the presence of a pseudogap causes the
pre-pairing of electrons so as to make the superconductivity
easier \cite{Ding:1996pso}. We have checked that this phenomenon
is always observed in the   small $k$ region under the condition
that the system falls into an insulating phase with non-zero
$\beta$, independent of the values of the other parameters. We
will further verify the connection between the formation of the hard
gap and the superconductivity in the next subsection.

Interestingly, the critical temperature with
vanishing $k$ is much higher than that saturated constant for
large $k$ which would dual to a metallic phase. This phenomenon
definitely deserves further investigation in future.

\item It is also interesting to plot $T_c$ as a function of
$\lambda$ with other parameters fixed (Fig.~\ref{fig-Tc}(b) and (c)). For large $k$, $T_c$
decreases monotonically with the increase of $\lambda$, as found in the simplest
superconductor model on the Q-lattice \cite{Ling:2014laa}. For
small $k$, however, $T_c$ rises as $\lambda$ increases! As we mentioned
in the previous section, large $\lambda$ always points to an
insulating phase, but the key difference is that a hard gap
presents in the small $k$ region. Furthermore, for a fixed $k$ with
small value, we find the larger $\beta$ makes the hard gap more
evident, such that the corresponding $T_c$ is higher.
\end{itemize}

Next we explicitly demonstrate that the formation of hard gap
indeed leads to a higher critical temperature of superconductivity,
by calculating the optical conductivity in the insulating phase.


\subsection{Critical temperature and gap in the normal state}

In this subsection we will visualize our above statement,
arguing that the presence of a hard gap in the insulating phase will
make the condensation easier.

The connection between the formation of the gap and the formation
of the condensation can be investigated by examining how critical
temperature behaves when the gap in optical conductivity emerges
with the variation of certain parameters. In what follows we fix
$b_0 = 0.5$, {\it i.e.}, the chemical potential dual to the
Maxwell field $B$, which means that we essentially work in the grand
canonical ensemble. After that, we examine explicitly the relation
between the gap and the condensation when changing system
parameters $\beta,\,\lambda,$ and $\,k$.

The optical conductivity of dual quantum field theory can
be calculated by a linear perturbation theory in bulk geometry.
To this end, we turn on the following self-consistent perturbations
\begin{equation}\label{perturbation}
  \delta A_x = a_x (z)e^{-i\omega t}, \; \delta B_x = b_x(z)e^{-i\omega t}, \;\delta g_{tx}=h_{tx}(z)e^{-i\omega t},\;\delta \Phi=ie^{ikx}z^{3-\Delta}\varphi(z)e^{-i\omega t}.
\end{equation}
Once the background solution is obtained, we can numerically solve the corresponding linearized perturbation equations with variables
$(a_x(z),b_x(z),h_{tx}(z),\varphi(z))$ with the ingoing boundary conditions at the horizon, and
read off the optical conductivity in response to the $B$ field along
the $x$-direction in terms of
\begin{equation}\label{cond}
\sigma(\omega)=\left.\frac{\partial_z b_x(z)}{i\omega b_x (z)}\right|_{z=0}.
\end{equation}
There are some key points to consider in the analysis:
\begin{itemize}
  \item We are only interested in the electrical response to the Maxwell field $B$, therefore on the boundary only $b_x(0)$ is turned on, with the other gauge field perturbation $a_x(0)$ set to $0$.
\item To ensure that we are extracting the current-current correlator, the perturbations are required to satisfy an additional boundary condition $\varphi(0)-ik\lambda h_{tx}(0)/\omega =0$, obtained from diffeomorphism and gauge transformation \cite{Donos:2013eha}.
\end{itemize}

\begin{figure}
  \centering
  \newcommand{\picwid}{6cm}
  \includegraphics[width=\picwid]{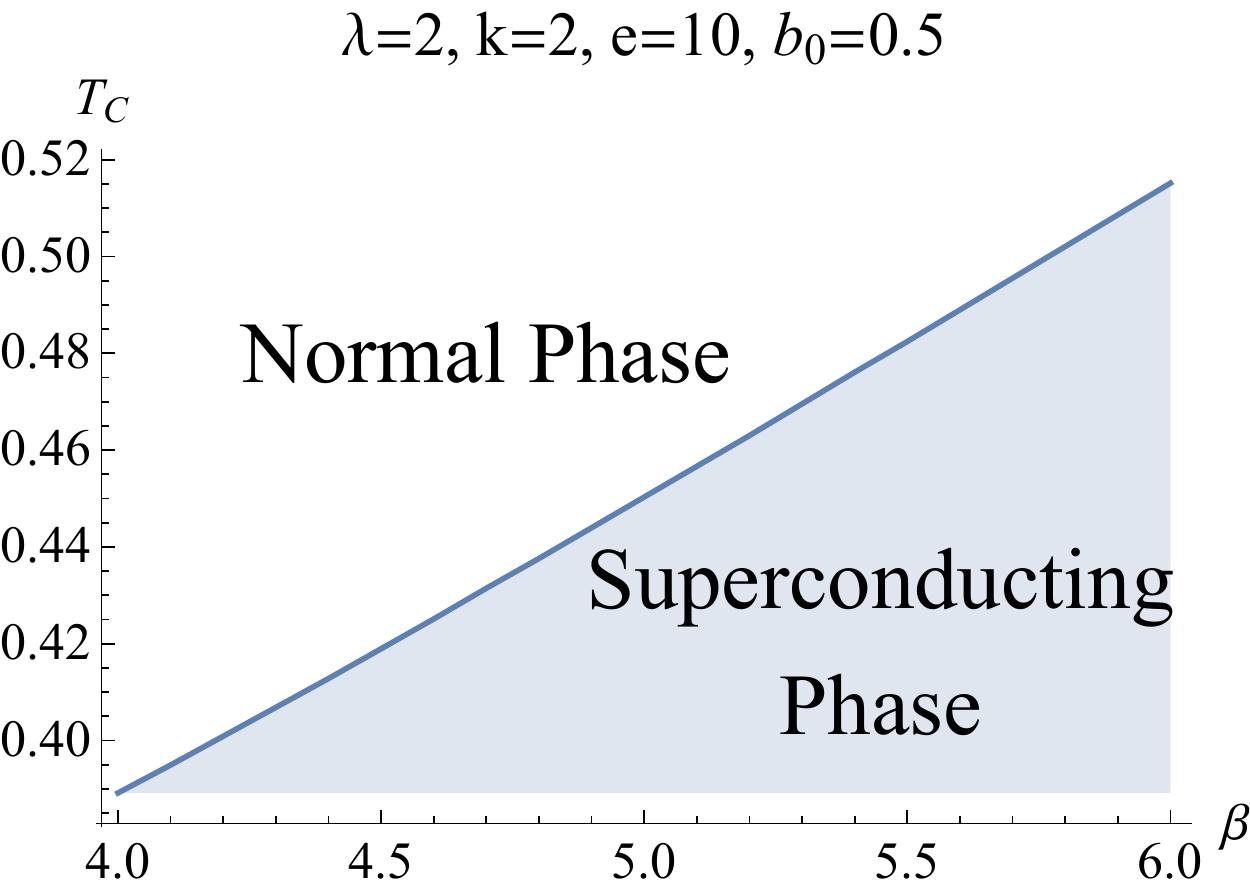} \qquad \includegraphics[width=\picwid]{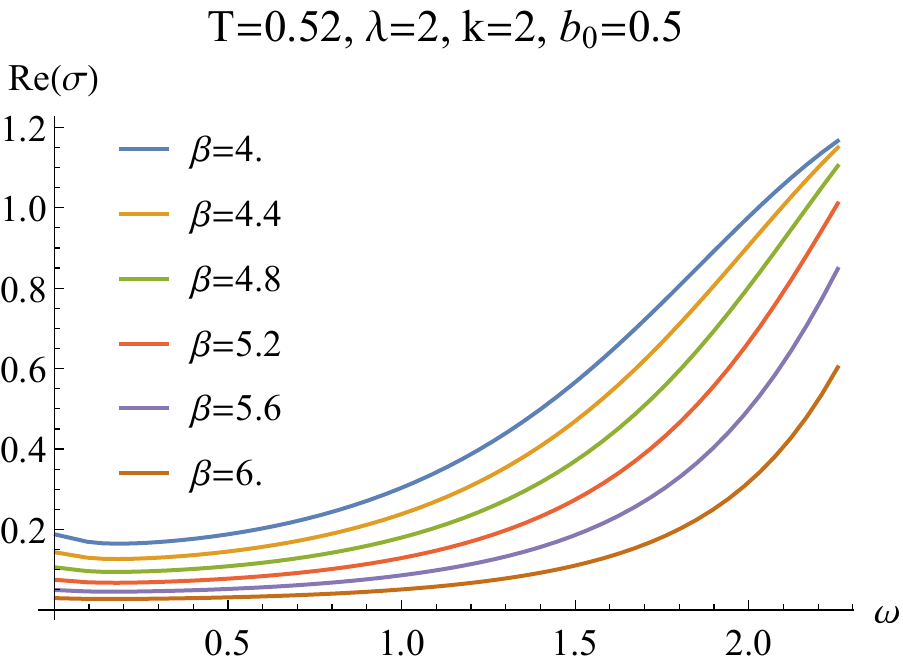}
  \includegraphics[width=\picwid]{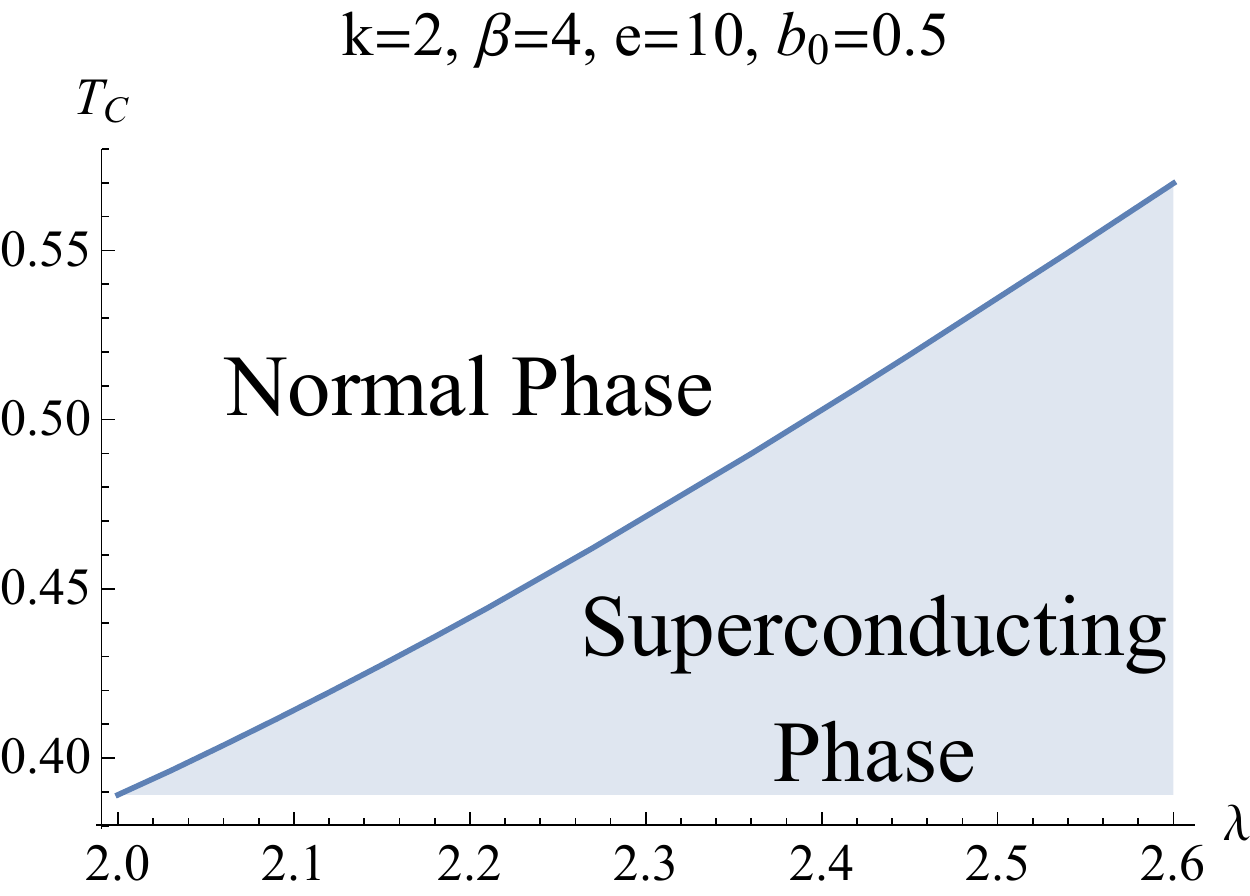} \qquad \includegraphics[width=\picwid]{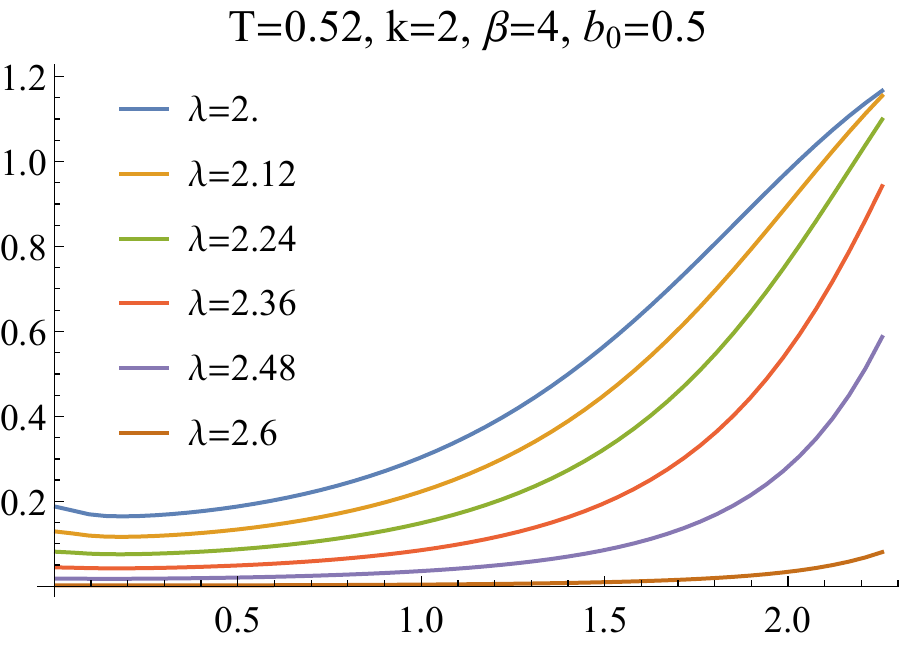}
  \includegraphics[width=\picwid]{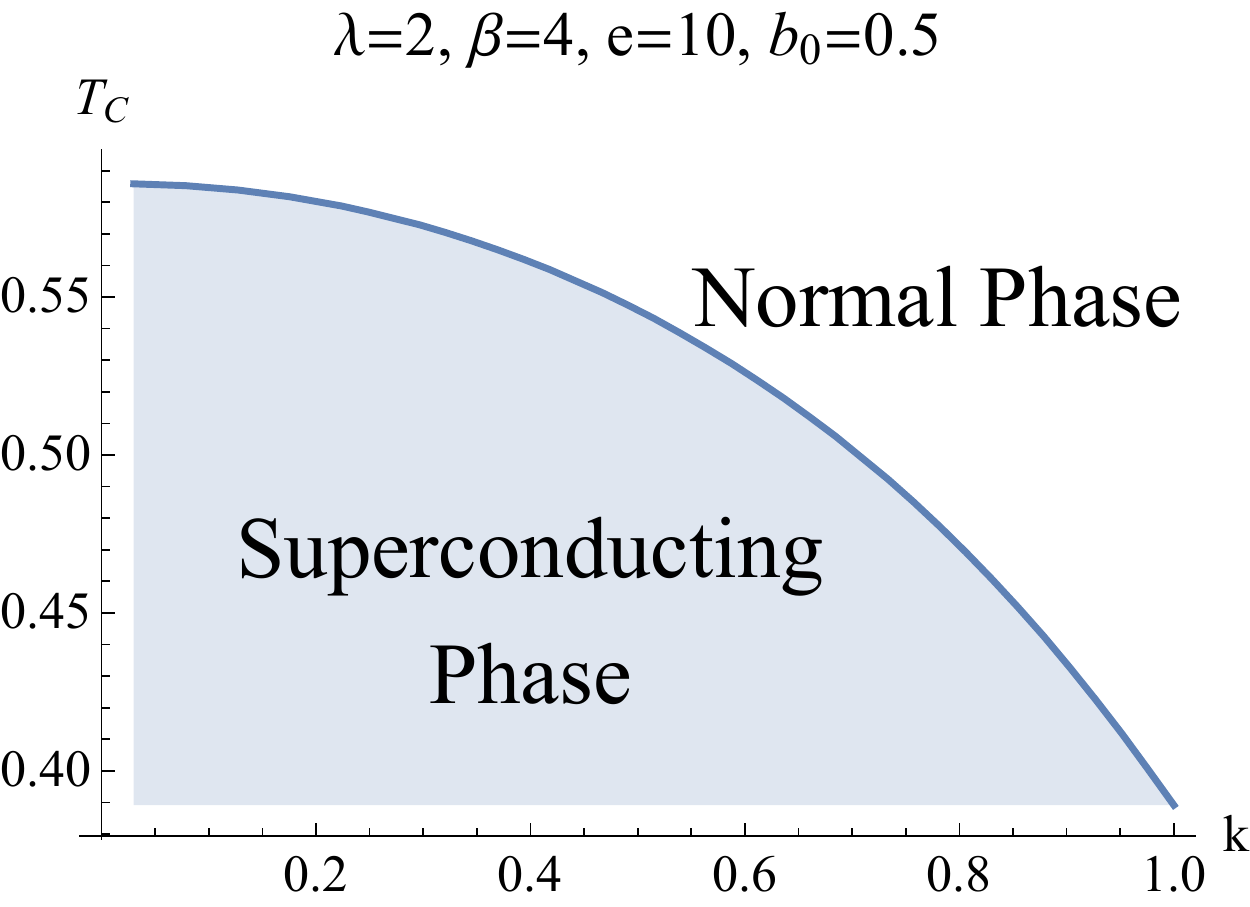} \qquad \includegraphics[width=\picwid]{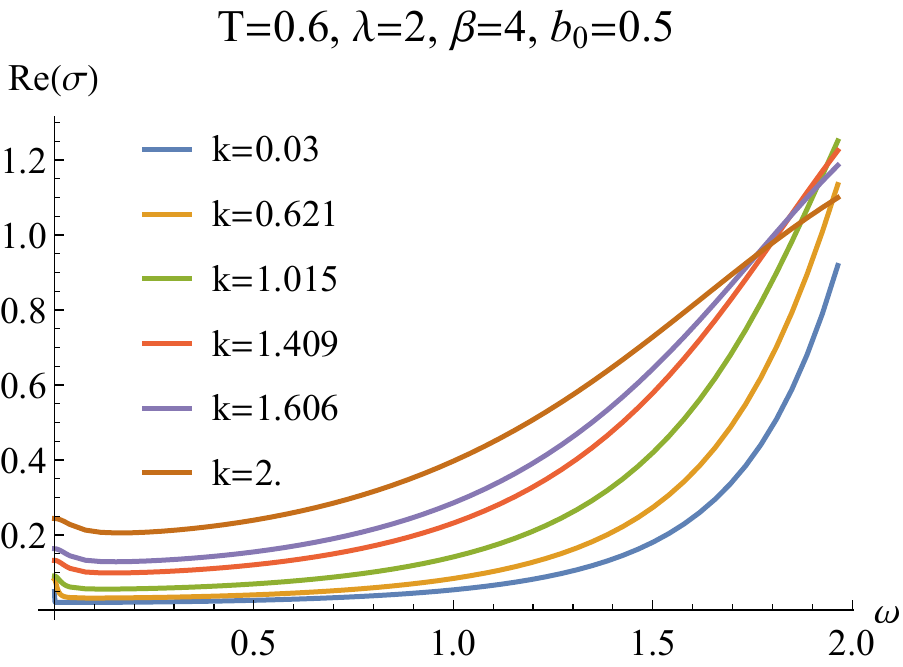}
  \caption{(a) The phase diagram for parameter $\beta$; (b) $\text{Re}\,\sigma(\omega)$ as a function of $\omega$ for several parameters of $\beta$; (c) the phase diagram for parameter $\lambda$; (d) $\text{Re}\,\sigma(\omega)$ as a function of $\omega$ for several parameters of $\lambda$; (e) the phase diagram for parameter $k$; (f) $\text{Re}\,\sigma(\omega)$ as a function of $\omega$ for several parameters of $k$.}
  \label{phasevsgap}
\end{figure}

We demonstrate the numerical results of the phase diagram and the
optical conductivity in Fig.~\ref{phasevsgap}. Figure~\ref{phasevsgap}(a) and (b) show that $T_c$ increases with
$\beta$, while the gap in optical conductivity becomes more
evident as $\beta$ increases. When varying parameter
$\lambda$, similar phenomena are observed, as can be seen from Fig.~\ref{phasevsgap}(c) and (d). Finally, $T_c$ increases with the decrease
of $k$, and indeed the gap becomes more pronounced when $k$
decreases, as shown in Fig.~\ref{phasevsgap}(e) and (f).
To sum up, $T_c$ increases when the gap becomes more evident. The
gap in optical conductivity resembles the role of the pseudogap in
the cuprates phase diagram.

We conclude that the presence of the hard gap in the
insulating phase makes the condensation easier, indicating that at
least in a holographic regime this kind of insulating phase with a
hard gap is qualitatively different from all the insulating
phases without a hard gap, as found in previous studies.

\subsection{Condensation}

\begin{figure}
\newcommand{\widoc}{7cm}
    \center{
    \includegraphics[width=\widoc]{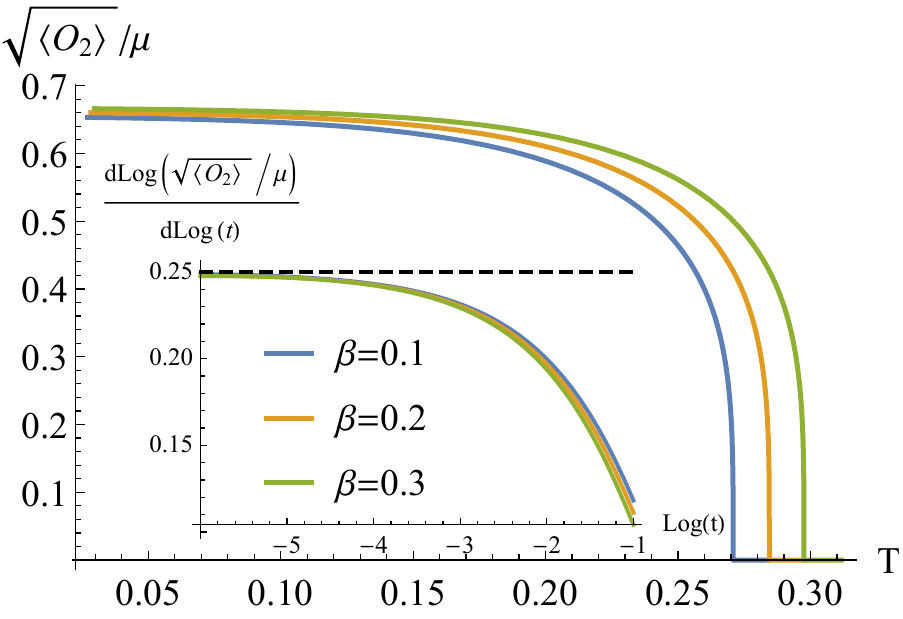}\quad
    \includegraphics[width=\widoc]{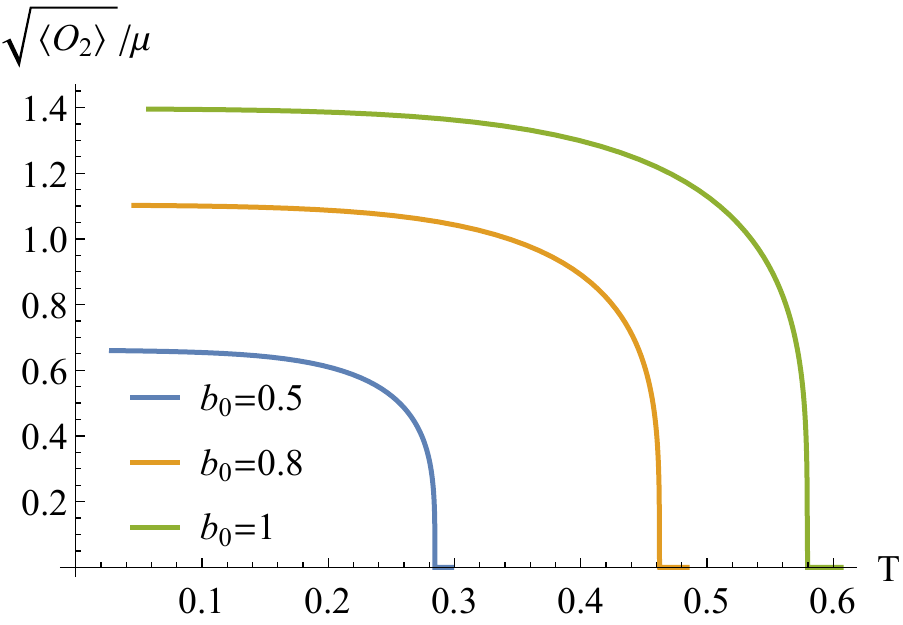}
    \caption{\label{fig-con}(a) The value of the condensation as a function of temperature
    with the given parameters $\lambda=2$, $k=0.03$, $b_0=0.5$, $e=10$ and several values of $\beta$.
    The inset is $\frac{d\log(\sqrt{O_2}/\mu)}{d\log(t)}\,\text{vs}\,\log(t)$ and the black dashed line is $\frac{d\log(\sqrt{O_2}/\mu)}{d\log(t)}=0.25$.
    (b) The value of the condensation as a function of temperature with the given parameters $\lambda=2$, $k=0.03$,
$\beta=0.2$, $e=10$ and several values of $b_0$.}}
\end{figure}

So far, all the above discussions on the conditions of
condensation are just based on the solutions to the eigen-value
problem of $e^2$ as described in Eq.~(\ref{eom-psi-eigen}), where each
background is fixed without the condensation of scalar hair. Such
an approximation is good enough for us to estimate the critical
temperature $T_c$. Now to explicitly construct a background with
scalar hair, we need to directly solve the coupled EOMs
(\ref{eom-g})-(\ref{eom-psi}) with the ansatz (\ref{bs}), which
can be done numerically with the standard spectral
method. At the end of this section we demonstrate a result for the
condensation of scalar hair $\sqrt{\langle O_2\rangle}/T_c$ as a
function of the temperature $T/T_c$ in Fig.\ref{fig-con}.

In conventional BCS theory, the superconductivity phase transition
is second order, which is described by $\sqrt{\langle O\rangle}\sim
t^{1/4}$ where $t\equiv (1-T/T_c)$. It is interesting to examine
the order of the phase transitions demonstrated in our model. As a
representative example, we show
$\frac{d\log(\sqrt{O_2}/\mu)}{d\log(t)}\,\text{vs}\,\log(t)$ as
the inset in Fig.~\ref{fig-con}(a). It is
readily seen that the superconductivity phase transition in our
model is also second order, since
$\frac{d\log(\sqrt{O_2}/\mu)}{d\log(t)}\sim 0.25$ uniformly.

\subsection{Optical conductivity in the superconducting phase}\label{sec-spt}

In this subsection we investigate the frequency
dependent behavior of the optical conductivity in the superconducting
phase, with a focus on the influence of the hard gap in the insulating
phase before the condensation takes place.

\begin{figure}
\centering
\newcommand{\hcond}{5cm}
\includegraphics[height=\hcond]{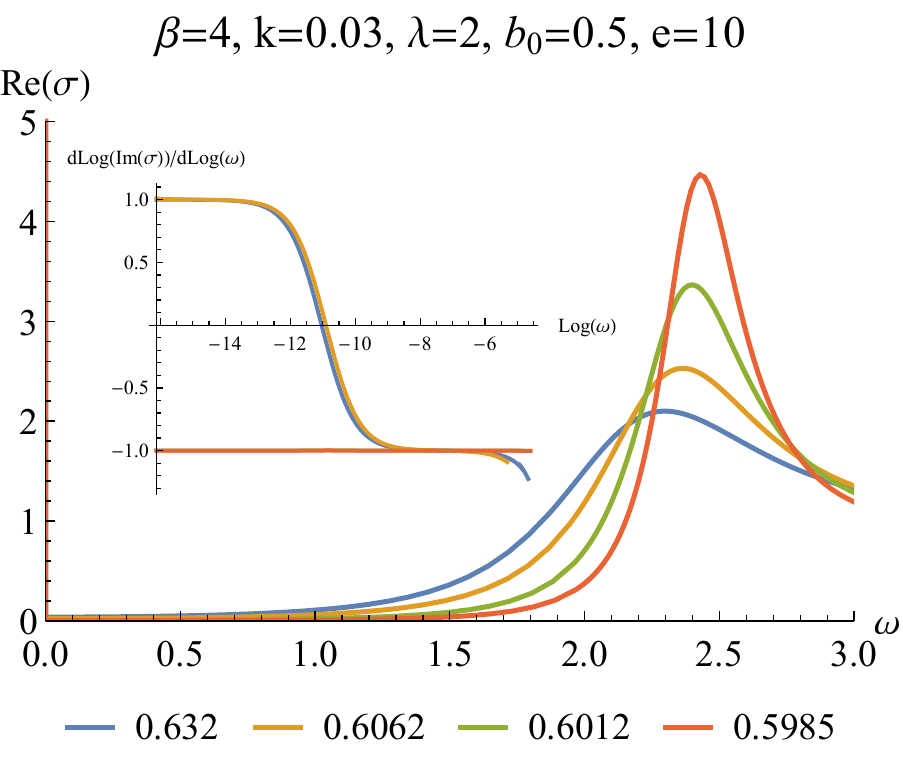}\qquad
\includegraphics[height=\hcond]{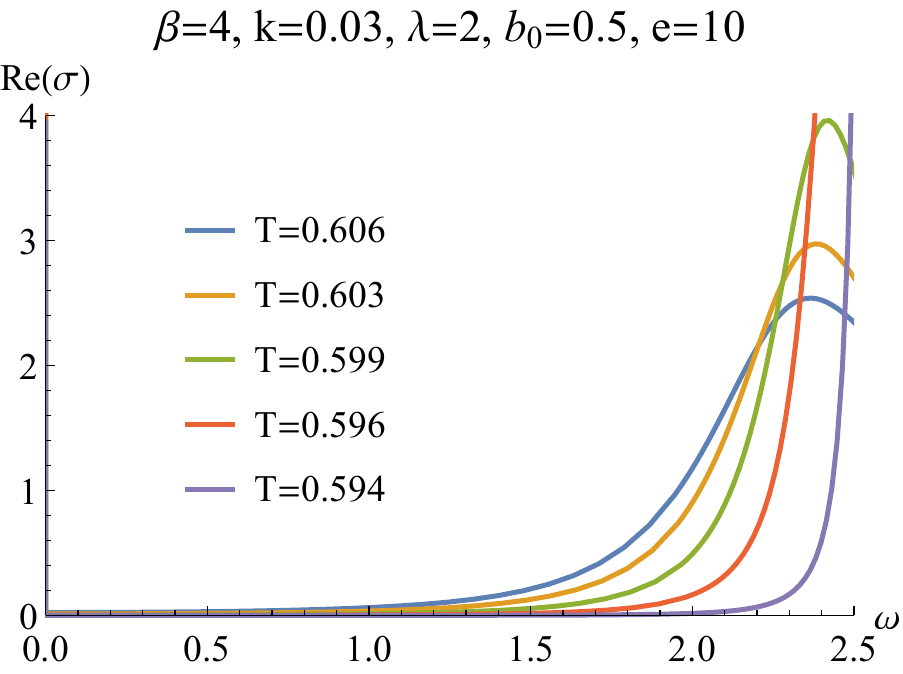}
\caption{(a) $\text{Re}\sigma(\omega)$ at four different temperatures
specified by the plot legend. The inset is $d\log \text{Im}
\sigma(\omega)/d\log (\omega)$. (b) $\text{Re}\,\sigma(\omega)$ at several temperatures below $T_c$.}\label{condphase}
\end{figure}

First, we illustrate the process of phase transition
with the behavior of the optical conductivity. When the system
transits from a normal phase to a superconducting phase, the DC
conductivity changes from a finite value into a $\delta$ function.
Thanks to the lattice structure which breaks the translational
invariance in our formalism, this process now can be observed
manifestly in optical conductivity. Nevertheless, a divergent DC
conductivity cannot be captured by taking the limit $\omega \to
0$. Instead, the $\delta$ function can be reflected by $\text{Im}
\, \sigma(\omega) \sim \omega^{-1}$ in the low frequency region,
in light of the Kramers-Kronig relations. Figure~\ref{condphase}(a) demonstrates $\sigma(\omega)$ at four
different temperatures across the phase transitions. The critical
temperature $T_c \simeq 0.606$ at parameter $\beta =4, \, k =
0.03, \, \lambda =2, \, b_0 = 0.5, \, e=10$. For insulating
phases with $T=0.632,\, 0.6062$, which are above $T_c$, it is found
that in the low frequency region $d\log
\text{Im}\,\sigma(\omega)/d\log (\omega)\sim 1$. This is expected
since the imaginary part of Drude conductivity $\text{Im}\, \sigma
\sim \omega$ \footnote{For both metallic and insulating phases,
the coherent and incoherent behavior of the conductivity have been
discussed in Ref.~\cite{Ling:2015exa}. For an insulating phase the
zero-frequency behavior of conductivity is not exactly Drude, but this linear relation with $\omega$ still holds.}.
For $T=0.0612,\, 0.5985$, which are below $T_c$,
$d\log \text{Im}\,\sigma(\omega)/d\log (\omega)\sim -1$, i.e. $\text{Im} \, \sigma(\omega) \sim \omega^{-1}$. Therefore,
the different scaling behavior of $\text{Im}\,\sigma$ clearly
demonstrates the process of phase transition.

Second, we are interested in how the gap behaves with respect to the
temperature below $T_c$. It is well-known in BCS theory, as well
as in many holographic models, that the gap becomes more and more
pronounced when the temperature drops. Figure~\ref{condphase}(b) shows the real part of optical
conductivity at several temperatures below $T_c$. It is easily
seen that the gap becomes more and more evident when the
temperature drops. This phenomenon is in agreement with
both BCS theories and many other holographic superconductivity
models.

\section{Discussion}\label{sec-dis}

In this paper we have constructed a novel holographic
superconductor based on a gravity theory with Q-lattices, in which
an interacting term characterized by parameter $\beta$ is involved
such that a hard gap can be observed in the insulating phase. In
contrast to all the previous holographic models, in which the
insulating phase suppresses the condensation of the scalar hair
and the critical temperature becomes lower in comparison with that
in the metallic phase, we have found that with the
emergence of the hard gap, the insulating phase will benefit the
condensation and the critical temperature becomes higher whenever
the hard gap becomes more evident. The phase transition from an
insulating phase to a superconducting phase has been explicitly
justified by examining the behavior of the imaginary part of
optical conductivity. All above phenomena imply that the hard gap
in this model resembles the role of the pseudogap in cuprate phase
diagrams. Therefore, this work may provide valuable insights into
the mechanism of high $T_c$ cuprates from the holographic side.

We believe that the fact of the presence of the hard gap in the insulating
phase benefiting the superconductivity should be robust and
general in the holographic approach, in the sense that it does not
depend on the specific choice of parameter values in this model,
nor on the specific formalism of the setup. Basically, we think the
formation of a hard gap in the optical conductivity implies that
more electrons are pre-paired in the dual field theory, such that
the instability of the system can be induced at a higher
temperature, leading to the condensation of the scalar hair.

The current setup in our model has one weak point that should be
mentioned. As we have noticed and pointed out in
Ref.~\cite{Ling:2015exa}, the existence of background solutions is
strongly constrained by the values of the parameters. In particular,
whenever an insulating phase with a hard gap is achieved, the bulk
geometry dual to a superconducting phase can be obtained only in
some region below the critical temperature. Numerical solutions do
not exist when the temperature is decreased further. Although we
could adjust the charge $e$ as well as the other parameters to
have solutions in an arbitrarily low temperature region,
not all the parameters guarantee the existence of solutions
in the zero temperature limit.
Mathematically, it is not rare to
encounter a gravitational system without domain wall solutions in
certain ranges of parameters or boundary conditions \cite{Perlmutter:2010qu}. From the dual theory side,
when the hard gap is formed, the optical conductivity tends
to vanish in the low frequency region. Further increasing $\beta$ or decreasing the
temperature would lead to negative conductivity, which is
of course unphysical. In our model this unphysical
phenomenon is not encountered due to the
absence of solutions to the Einstein equations for these parameters. It is still
desirable to obtain a system solvable for
any given parameters and which exhibits all the interesting
behaviors revealed in our current work.
We expect to improve this by adjusting the setup in the holographic models.

Along this direction, there is much worthy of further
investigation. One avenue is to investigate the spectral function
of a probe such as holographic fermions to see if the two
fundamental features of Mott insulator could be observed, namely,
the hard gap and the spectral weight transfer, and then observe
its variations during the superconducting phase transition. Since
the fermionic field only feels the bulk geometry as a probe, its
spectral function would be very sensitive to the specific
couplings of the fermion and background. We hope to explore this
issue in the near future.


\centerline{\rule{80mm}{0.1pt}}

\end{document}